# 2D full-band, Atomistic Quantum transport in L-shaped Vertical InSb/InAsn-TFETs

Bhupesh Bishnoi[1] and Bahniman Ghosh[1,2]

[1]DEPARTMENT OF ELECTRICAL ENGINEERING, INDIAN INSTITUTE OF TECHNOLOGY, KANPUR, 208016, INDIA

Email: bbishnoi@iitk.ac.in

[2]MICROELECTRONICS RESEARCH CENTER, 10100, BURNET ROAD, BLDG. 160, UNIVERSITY OF TEXAS AT AUSTIN, AUSTIN, TX, 78758, USA

Email: bghosh@utexas.edu

ABSTRACT

In the present work, we have investigated the performances of L-shaped Vertical broken bandgapheterostructureInSb/InAsn-channel tunnel field-effect transistors (TFETs) of 4 nm thin channel structures with the gate lengths of 20nm. We have used a 3-D, full-band, quantum mechanical simulator based on atomistic $sp^3d^5s^*$spin-orbital coupled tight-binding method.In this L-shaped nonlinear geometry the gate electric field and tunnel junction internal field are oriented in same direction. A broken narrow bandgap (BG) structure has another advantage that transport is by mixture of electrons/holes. TFETs are promising devices for lowpower logic design due to low subthreshold swing (SS)and high $I_{on}/I_{off}$ ratio. We investigate current–voltage characteristics, ON-current, OFF-current andsubthreshold swing as function of equivalent oxide thickness, gate length, drain length, gate undercut, High-K, and drain thicknessfor L-shaped nonlinear geometry tunnel FET for low subthreshold swing and lowvoltage operation. To study 2D electronic transport in this L-shaped nonlinear geometry we used Non-Equilibrium Green Function (NEGF)based quantum transport method using $sp^3d^5s^*$tight binding model in which Poisson-Schrödinger solver self-consistently iterates to obtain potentials and Local Density of States (LDOS). The advantage of this method is that it can handle arbitrary geometries and complicated 2D structures like Band-to-Band tunnelling (BTBT) devices. NEGF quantum transport method gives output in terms of Poisson potential and space charge in 3D, energy resolved transmission and spectral function and Local Density of States of electrons/ holes in space and energy.

## INTRODUCTION

In the current era of nanoelectronic devices,leakage power dissipation is a fundamental bottleneck. At room temperature,at least 60 mV of gate voltage is required in today's MOS field-effect transistors (MOSFETs) to change current by an order of magnitude for switching operation. But, the supply voltagescalingof last three decadesreduces the energy window needed for transistor switching.In present scenario Tunnel field-effect transistors (TFETs) are promising candidates due to their steep subthreshold swing, better ON to OFF current ratio and high drive current at low voltage operation. Hence, TFETs will reduce the overall power consumption of nanoelectronics integrated circuits. [1]TFETs are able to break the fundamental limit of subthreshold swing (SS)of 60 mV/decade because in TFETs charge carriers are injected into the channel by quantum-mechanical band-to-band tunneling (BTBT) of valence band electronsinstead of thermionic emission process of conventional MOSFETs. Hence, SS can go below 60 mV/ decade.[2] However, Tunnel field-effect transistors (TFETs) have low $I_{ON}$ current, which reduces speed response of TFETs based circuit. But, carefully designed electrostatically optimized device structure can improve $I_{ON}$ current.As per the ITRS roadmap the future target parameters for TFETs are: $V_{DD}$ is less than 0.5 V, 100milliamperes of $I_{ON}$,$I_{ON}/I_{OFF}>10^5$ and SSfar below 60mV per decade. A steep slope and high tunnelling current can be achieved if minute change in gate voltage will change source tunnelling barrier's transmission probability from zero to unity. [3]To achieve this, instead of lateral TFET geometries,vertical transistor geometries are proposed to enable lower off-state leakage and high drive current with better gate electrostatics control. In vertical transistor geometries the tunnel junction is oriented such that tunnel junction internal field and gate field are aligned with each other. [4]To increase the $I_{ON}$ and hence the tunnelling probability one seeks efficient ways to reduce the bandgap. Materials with broken-gap (BG) band alignment e.g. InSb/InAsare promising to build TFETs because of a zero band overlap and narrow band gap of 0.17 eVand 0.36 eVfor InSb andInAs, respectively.[5]The combination of InSb and InAs material gives rise to type-III brokenbandgap alignment (i.e., conduction band minimum of one material is lower in energy compared to the valence band maximum of other) which inherently favors tunnelling and leads to low resistivity in the junction. Moreover lightereffective masses also enhance the probability of tunnelling for InAs andInSb. [6]

In theheterostructure TFETs, the materials are chosen such that at one side of junction one has a smallerbandgap material and hence width of the energy barrier at that side of junctioncan be decreased in the ON-State while on the other side of the junction, the material has a largerbandgap.Hence this creates the energy barrier width of largest possible size at that side of junction which keeps the off state current low.Materialsfrom III–V group are quite attractive as they allow different band-edge alignments and have smallertunnellingmass.Group III–V materials have another advantage of lattice-matched growth during growth processing.[3]In this article, we demonstrate various electrostatic and geometrical considerations that influence the scaling and design of vertical L-shaped InSb/InAsheterojunction TFETs as demonstrated experimentally.[7,8]In the vertical L-shaped geometry tunnelling barrier width is optimized by applying the gate modulated electric field as the electric field and tunnelling path are aligned in the structure.Hence, in the design we have to overlap the gate on the tunnelling region as source region is covered by the top gate. By this design practice a factor of 10 improvements is visible in ON-current and steep subthreshold swing can be achieved.[9]In commercial two dimensional device simulators like Synopsys TCAD, tunnelling models are used based on Wentzel-Kramers-Brillouin (WKB) approximation. These models are1-D in nature and neglect confinement effects and band quantization effects. Since in our device structure, the electrostatics and current path are 2-D, for more accurate analysis, we must go beyond the 1-D tunnelling models [10]. For that purpose we used 3-D, full-band, quantum mechanical simulator based on atomistic $sp^3d^5s^*$spin-orbital coupled tight-binding method.We investigated the performances of L-shaped Vertical broken bandgapheterostructureInSb/InAsn-channel tunnel field-effect transistors (TFETs) of 4 nm thin channel structures with the gate lengths of 20 nm.A full self-consistent quantum mechanical simulation including electron-phonon scattering can in principle describe TFETs with high accuracy. However, it requires an extremely high computational intensity to solve such NEGF equations [11]. Usually coherent transport simulations are performed to obtain the upper device performance limit. [5]

Historically, the evolution of tunnelling device is first proposed by Quinn *et al.* in 1978 which is the gated p-i-n structure.[3]Intensive research work is going on in TFETs as power-supply scaling below 0.5 V is possible in these devices and at low voltages TFETs can outperform aggressively scaled MOSFETs. [12, 13]Recently, a new vertical geometry TFET has been experimentally demonstrated. [7,8]Steep subthreshold swing can be achieved by in line field orientation of Gate field and tunnel junction internal field as analytically formulated by Zhang *et. al.*[14] On the basis of same concept, Hu *et. al.*proposed a new geometric configuration of pocket TFETs and Salahuddinand Ganapathialso proposed similar TFETs [15, 16].Ford *et.al.*experimentally realized these devices [17].Asra*et.al.*reported the VerticalSi homojunction TFETs [18] and Agarwal*et al.*reported the InAshomojunction TFETs. [19]

## II. DEVICE STRUCTURE

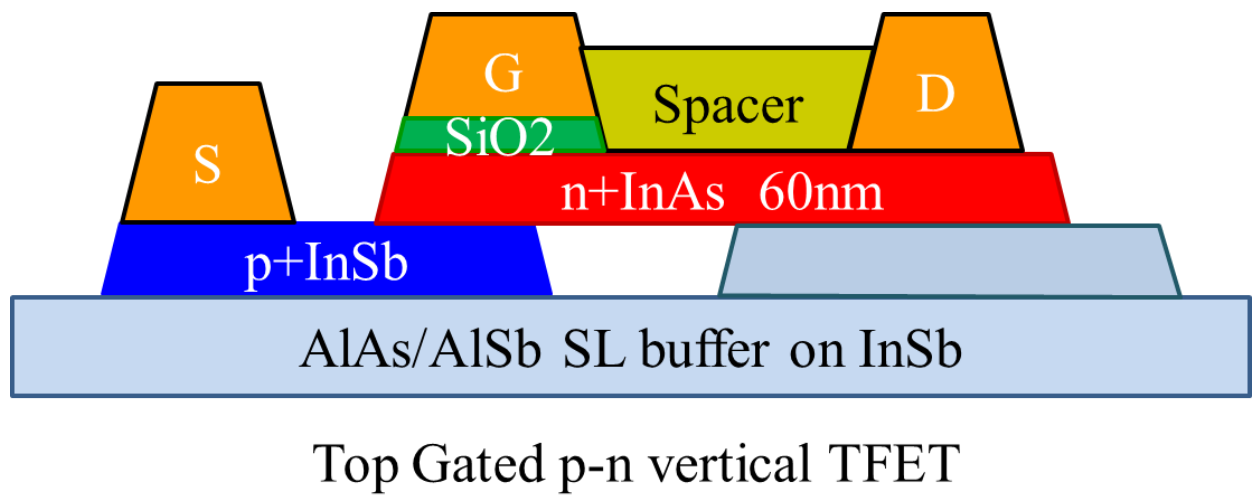

Fig.1. Cross section cartoon diagram of the simulated InSb/InAs TFET device

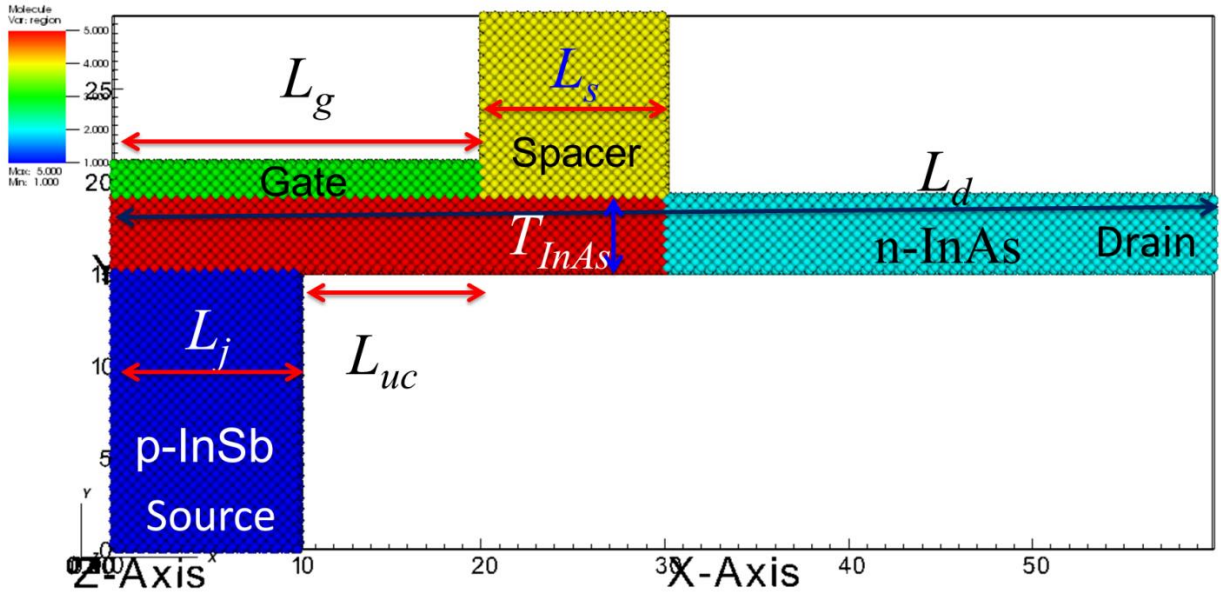

Fig.2. Atomic Structure and Geometry of L-shaped InSb/InAs TFET

In figure 1we have shown cross section cartoon diagram of simulated InSb/InAsTFET device structures. The heterostructureInSb/InAs TFET consists of 4 nm n-type InAs channel with a doping density of $5\times10^{17}$ cm$^{-3}$ and a 10 nm p-type InSb source injector with a doping density of $4\times10^{18}$ cm$^{-3}$. N-typeInAs has a band gap of 0.36 eV and p-type InSb has a band gap of 0.17 eV. The source and drain regions measure 10nm and 60nm in length, respectively. The drain thickness $T_{InAs}$is set to 4 nm. A SiO2 gate thickness of 1.9 nm is used initially and then varied to 4 nm. The gate metal length defined the gate length ($L_G$) of device.To fulfill the ITRS requirement we started our simulation with 20 nm gate length ($L_G$). We also used HfO$_2$ high-Kdielectric as gate material for another study. Between the drain and gate contact SiO$_2$ spacer of an overlap length ($L_S$) is used. Overlap spacer decoupled the gate–drain region and reduced the ambipolar conduction. The drain length ($L_D$) is defined by the length of the drain and junction length ($L_J$) is the length of active tunnellingjunction.Additionally the InAs channel has an undercut of a length ($L_{UC}$) which is necessary to achieve a steep subthreshold swing. Hence gate overlaps the channel. Type-III (broken gap, BG) band alignment of L-shaped InSb/InAsTFETs is shown infigure 3.Direction of transport in the channel is along <100> crystal axis and orientation of surface is along (100).Due to quantization effect in junction length ($L_J$) smaller than 10 nm BG characteristicmay vanish at BG heterostructure. However, we can shift the overall bandstructure by the tight-binding onsite energy to obtain different band alignments [20]

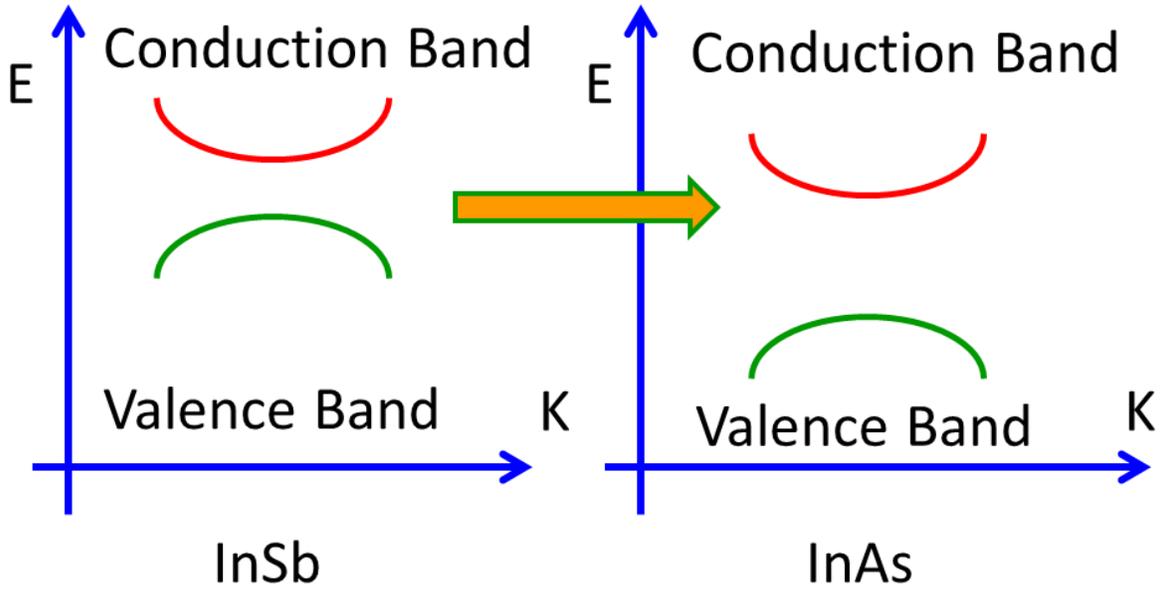

Fig.3. Type-III (broken gap) band alignment of L-shaped InSb/InAsTFET

SIMULATION APPROACH

We have used a 3-D, full-band, quantum mechanical simulator based on atomistic $sp^3d^5s*$spin-orbital coupled tight-binding method which solves Schrödinger and Poisson equations self-consistently.In the present work, a method combining the semiclassical density and Non-equilibrium Green's Function (NEGF) formalism is used to achieve an efficient simulation of InSb/InAsTFETs. This model assumes an equilibrium carrier distribution. States are filled according to quasi-Fermi levels, which could vary spatially and mimic strong scattering. By using Eq.1 the triangular quantum well could be populated, similar to the case of dissipative transport.

$$n = N_c \frac{2}{\sqrt{\pi}} F_{\frac{1}{2}}(\eta_c) \qquad (1)$$

Taking one unit cell from lead and calculating the bandstructure in the transport direction, one can obtain the effective bandgap after confinement. The effective mass is calculated from the doping density ($N_D$) and the doping degeneracy ($\eta_c$) in the contacts according to Eq.1. The doping degeneracy is calculated using the $sp^3d^5s*$ tight-binding model self-consistently for the same unit cell of TFETs with equilibrium boundary condition. The modification of density of states due to confinement is included in the process. We used the quantum transport simulator which is multi-dimensional, massively parallel, based on atomistic $sp^3d^5s^*$spin-orbital coupled tight-binding representation of the band structure. Quantum transport simulation can be done either in the ballistic transport regime by Wave Function formalism or in diffusive transport regime with scattering mechanism by Non-equilibrium Green's Function formalism(NEGF). Wave Function (WF) formalism is computationally efficient but numerically identical to the Non-equilibrium Green's Function formalism(NEGF). [21] In WF formalism we are solving sparse linear systems of equations while in NEGF formalism we are solving matrix inversion problems. Our simulations run with 960 number of energy points and 31 number of momentum points along the transport path. In theactive region of device every atom is represented by a matrix.In our simulation number of active atoms taken into account in the Schrödinger equation is 8696. The overall Hamiltonian matrix of the system is the tri-diagonal block matrix of sparse blocks. Number of atoms in each atomic layer will decide size of the sparse blocks. Gate dielectric layer and spacer layer are modeled as imaginary materials layer which has infinite bandgap as they separate the gate contacts and InAs channel and do not participate in transport calculation. Hence in the Poisson equation they are characterized by their relative dielectric constants.

In the figure 4 flow of Quantum Transport simulation for L-shaped InSb/InAsTFETs is shown.

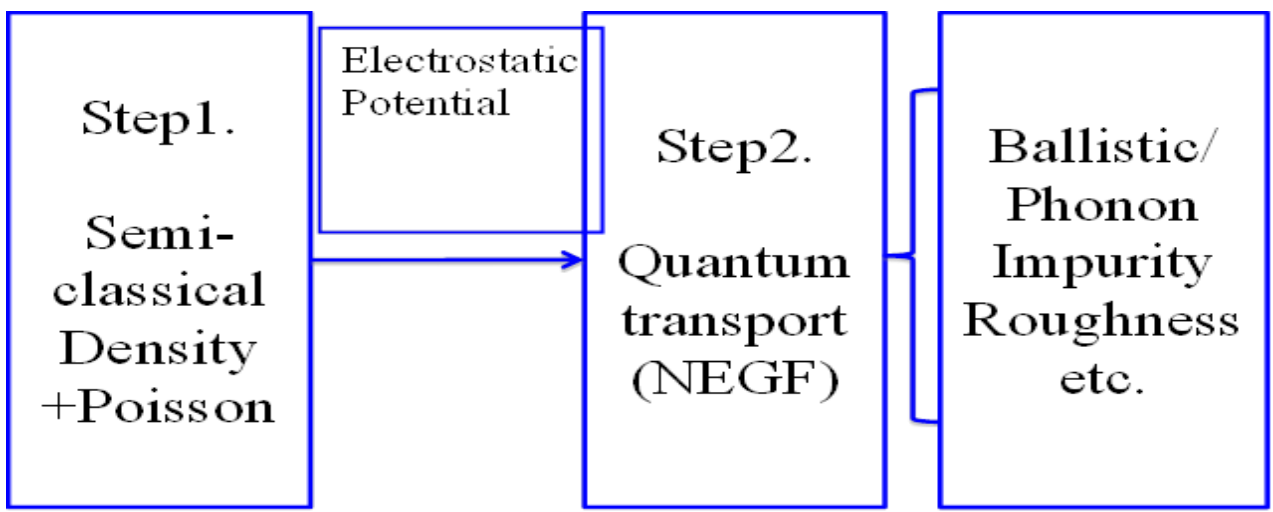

Fig.4. Flow of Quantum Transport simulation for L-shaped InSb/InAs TFET

In the device structures, holes and electrons are injected by the drain and source contacts with different wave vectors and energies and carrier charge densities and current densities are obtained. Electrons tunnel from valence band of p-InSb to conduction band of n-InAs and these tunneling electronscontribute to both charge densities and current.These carrier charge densities are self-consistently solved along with the calculation of electrostatic potential. Hence, the solutions of the Poisson and the Schrödinger equations are massively parallelized. [22]The band gap, effective masses and conduction and valence bands of various semiconductor materials are modeled with spin-orbit coupling. It also incorporates quantization effects due to narrow size. Neglecting these effects will increase the band gap and underestimate band-to-band tunnelling probability.Atomistic $sp^3d^5s^*$spin-orbital coupled tight-binding representation of energy band gap also accounts for the imaginary band dispersion. Hence, even in the absence of scattering process BTBT processes are accurately modeled for direct bandgap materials. [2-23] Inthe source quantization effects influenced the effective barrier height for tunneling and as a result density of states in the channel is changed. Hence, absolute current level changes.[9] Various local and nonlocal BTBT models are incorporated in the commercial device simulator but they do not address quantization effects.[12] For a hetero junction device no analytical model and theory exist for BTBT tunnelling in between an indirect and a direct semiconductor. [24]

RESULTS

A. energy-positionresolved local density of states LDOS(x,E) and energy-position resolved electron density spectrum $G_n(x,E)$ distribution

Figure 5 shows the energy-positionresolved local density of states LDOS (x,E) and energy-position resolved electron density spectrum $G_n(x,E)$ distribution of the 2-D BGInSb/InAsTFETsat $V_{DS}$= 0.3V ON-state conditionswith variation of $V_{GS}$from -0.1 V to 1.2 V with the step voltage of 0.1 V. In this ON-state biasing condition,gate modulates the position of the channel barrier which is due to the broken-gap energy band and hencechannel conduction band is pulled down below the source valence bandto increase the source injection.The ON-state is clearly visible in the Figure 5.It also shows the heterojunction broken bandgap at the source–channel interface.In the Figure 5 we also showedthe energy-positionresolved electron density spectrum $G_n(x,E)$ distribution which shows the occupation of LDOS (x, E) by the respective source and drain contact Fermi reservoirs. Energy-position resolved electron density spectrum is shown on log scale in the InSb/InAs TFETs at $V_{DS}= 0.3$ V. Figure 5 also shows original BG band alignment and amount of band shift in BG characteristic due to quantization.

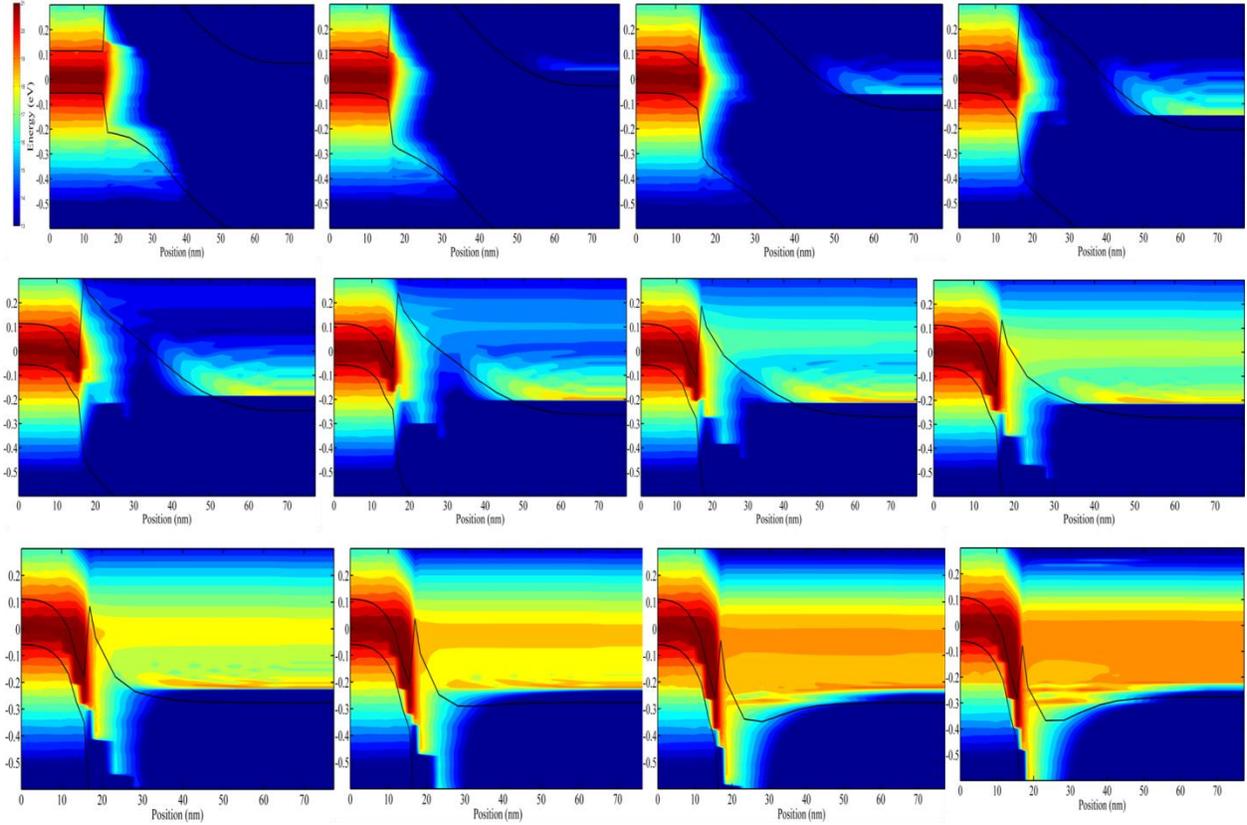

Fig. 5.Energy-positionresolved LDOS(x,E) and energy-positionresolved electron density spectrum $G_n(x, E)$ distribution in the ON-stateInSb/InAs TFET

Figure 6 shows the energy-position resolved local density of states LDOS (x, E) and energy-position resolved electron density spectrum $G_n(x, E)$ distribution of the 2-D BG InSb/InAsTFETsat $V_{DS}= 0.03$ V OFF-state conditions with variation of $V_{GS}$ from -0.1 V to 1.2 V with the step voltage of 0.1 V. OFF-state leakage current is mainly due to phonon absorption assisted tunnelling current. In the drain, carrier thermalization due to phonon emissions is also shown in the figure 6.

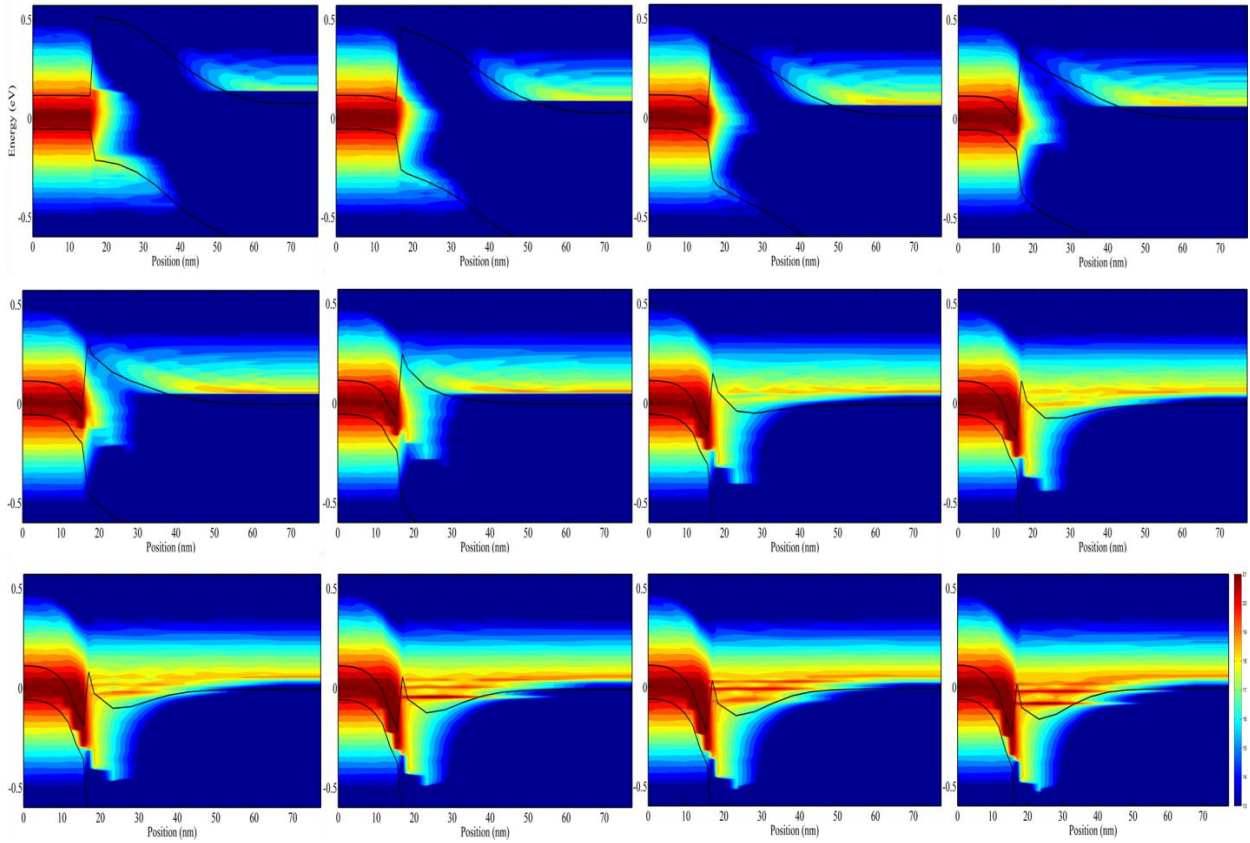

Fig. 6. Energy-position resolved LDOS(x, E) and energy-position resolved electron density spectrum $G_n(x, E)$ distribution in the OFF-state InSb/InAs TFET

B. Current-Voltage characteristics

Figure7 shows the $I_{DS}$-$V_{GS}$ transfer characteristics of InSb/InAsTFETs for different values of $V_{DS}$ of 0.3V and 0.03V. On applying Gate voltage $V_{GS}$ of 0.6 V for the on state drain voltage $V_{DS}$ of 0.3 V, an $I_{ON}$ current of 1 mA/μm, an $I_{ON}$/$I_{OFF}$ ratio of $10^{12}$ with subthreshold swing of about 21.73mV/decade are obtained. For both the cases, i.e., $V_{DS}$ = 0.3 V and 0.03 V, we obtain lowest value of drain current at same gate voltage, i.e., $V_{GS}$ = 0 V which indicates that the gate has good control on the tunnel junction. In the ON-state for gate voltage $V_{GS}$ higher than 1 V drain current almost saturates and the TFET gives high output resistance.

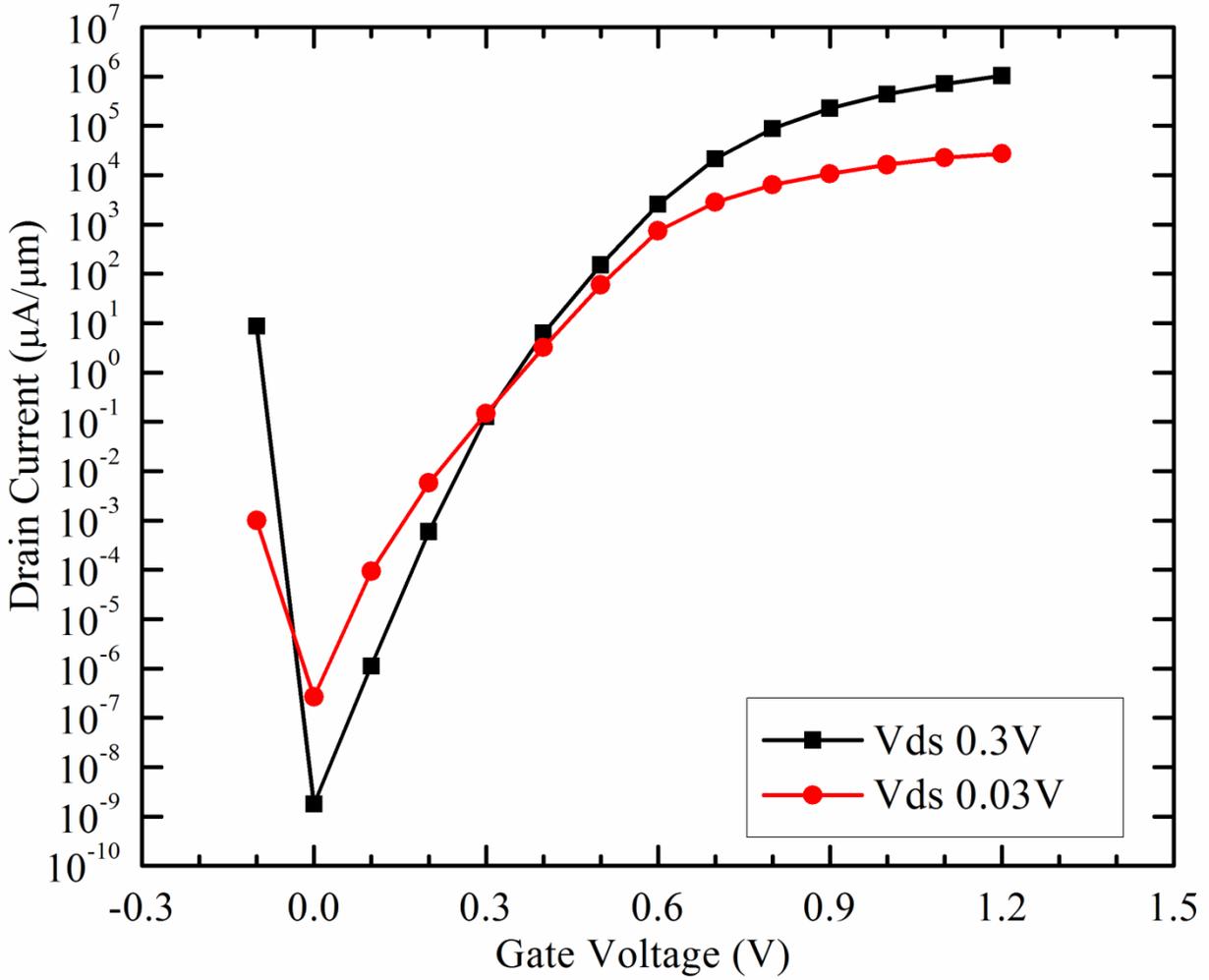

Fig. 7. $I_{DS}$–$V_{GS}$ characteristics of the InSb/InAs TFET at $V_{DS}$= 0.03V and $V_{DS}$= 0.3V

Figure 8 shows the $I_{DS}$–$V_{DS}$ output characteristics of BG InSb/InAsTFETs with Gate voltage variation $V_{GS}$ (0.1 V, 0.2 V, 0.3 V, 0.6 V, 0.9 V and 1.2 V). For $V_{GS}$ of 0.1V and below, the off current is mainly due to thermionic emission leakage and the current due to tunnelling process is negligible. In the ON state, withdrain voltage $V_{DS}$ of 0.3 V, the entire area of tunnelling junction is turned on and current density is uniform across the junction.Figure 9 shows the ON-State $I_{DS}$–$V_{GS}$ transfer characteristics of BGInSb/InAsTFETswith variation in temperature.Temperature dependence is found to be extremely minute from gate voltages of 0V to positive 1.2V, thereby clearly indicating direct band-to-band tunnelling.

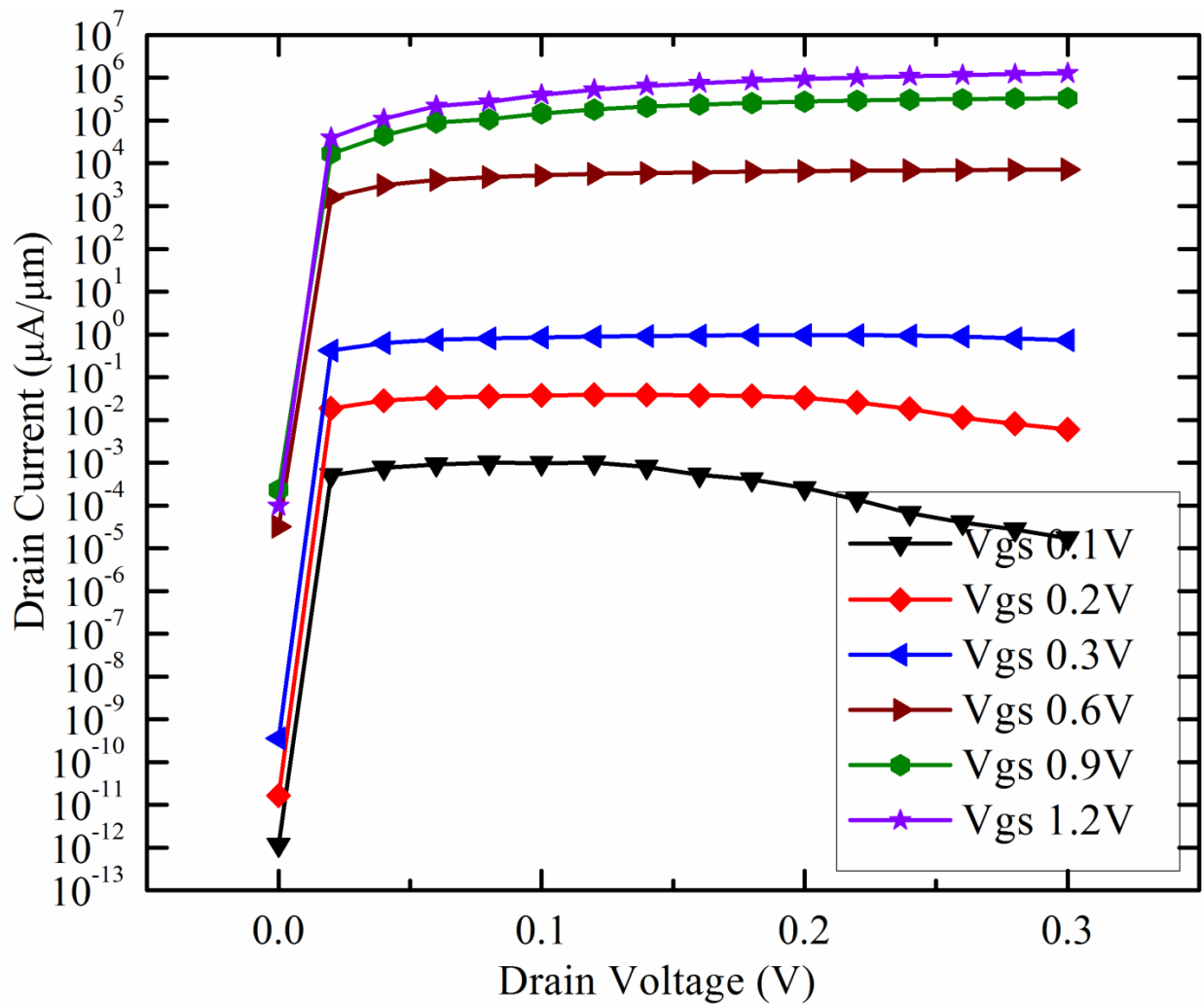

Fig. 8. $I_{DS}$–$V_{DS}$ output characteristics of BG InSb/InAs TFET with Gate voltage ($V_{GS}$) variation.

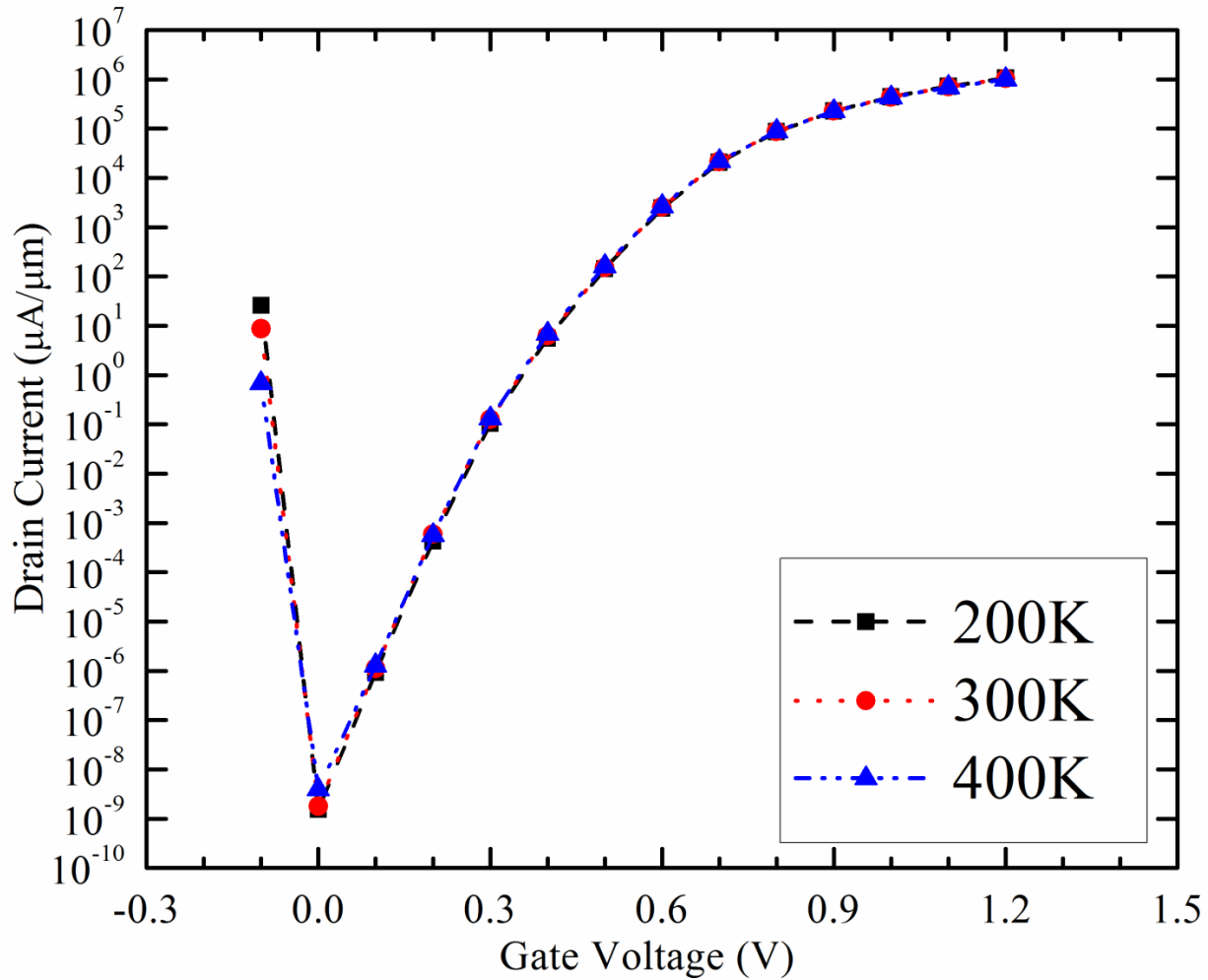

Fig. 9. $I_{DS}$–$V_{GS}$ characteristics of the InSb/InAs TFET at $V_{DS}$= 0.3V with variation in Temperature

C.    Effect of Geometry variation in InSb/InAs TFETs

We have investigated five different geometric variations. Figure 10 shows ON-state $I_{DS}$–$V_{GS}$ transfer characteristics of the InSb/InAs TFETs with variation in drain length $L_D$ extensions. The ambipolar current is estimated as in the real devices since the simulation includes the quantization effect and effective increase in channel bandgap. Electrostatic effect remains unaffected by the channel quantization and increase in channel bandgap. We observed that increasing the drain length ($L_D$) extension increases the ambipolar conduction current and hence turn-on characteristics and $I_{ON}$ current largely increases.

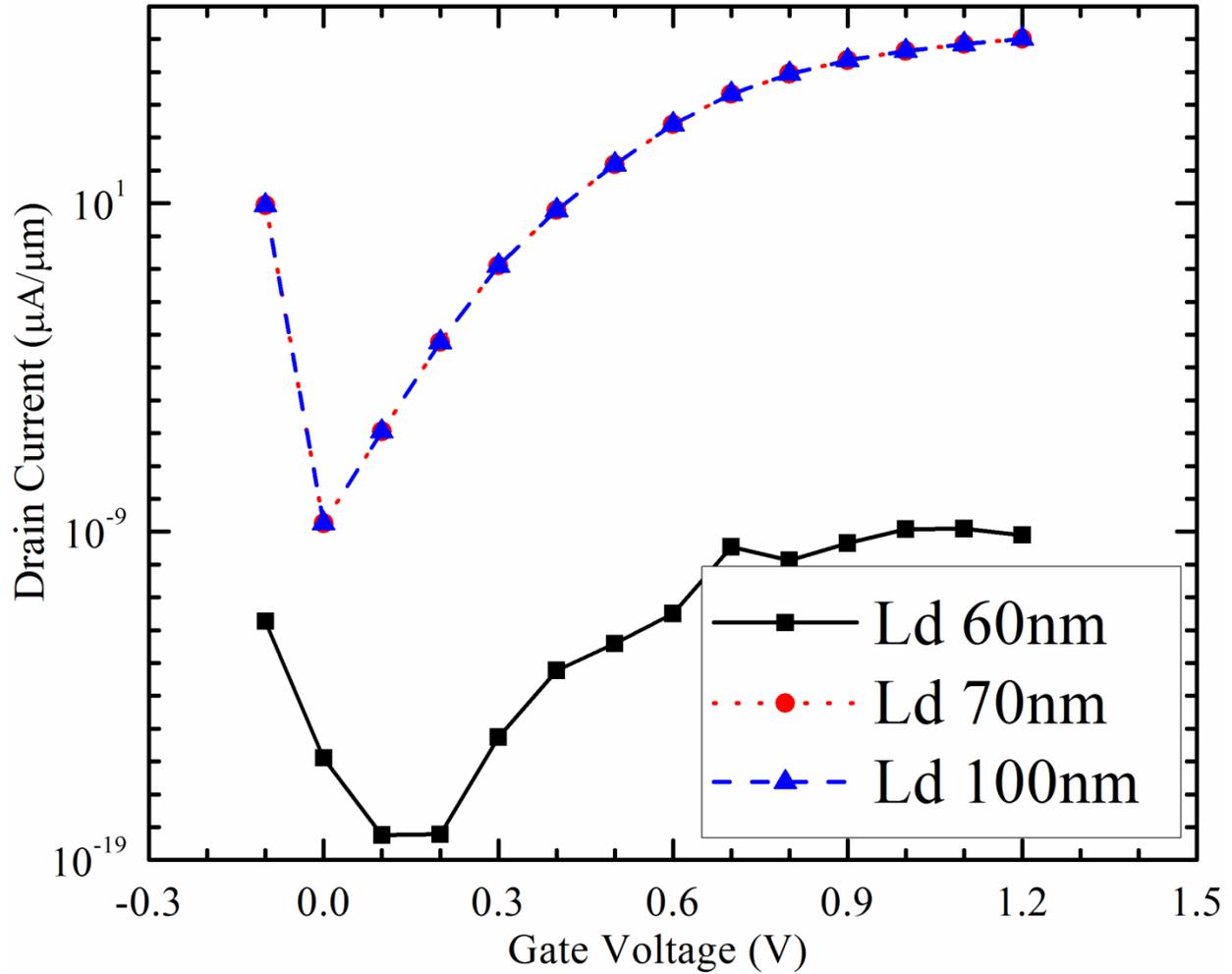

Fig. 10. $I_{DS}$–$V_{GS}$ characteristics of the InSb/InAs TFET at $V_{DS}$= 0.3V with variation in drain length ($L_D$)

Figure 11 shows the $I_{DS}$–$V_{GS}$ characteristics of the InSb/InAsTFETsat $V_{DS}$= 0.3 V with variation in undercut lengths ($L_{UC}$). For zero undercut lengths, $I_{ON}/I_{OFF}$ ratiois reduced by five orders of magnitudeand the subthresholdswing increases to 33.8 mV/decade. Subthreshold swing reduces to 21.73 mV/ decade and $I_{ON}/I_{OFF}$ ratio increases on increasing the undercut lengths to 10 nm. $I_{ON}$ current reduces on further increase in the undercut lengths as $I_{ON}$ current is proportional to the tunnel junction area and hence on increasing the undercut lengths $L_{UC}$ width-normalized $I_{ON}$ current density decreases. Here the optimization criterion is overlapping the gate on the tunnelling region. The scalability of InSb/InAsTFETs is limited by undercut lengths $L_{UC}$ which is necessary to achieve a steep slope.

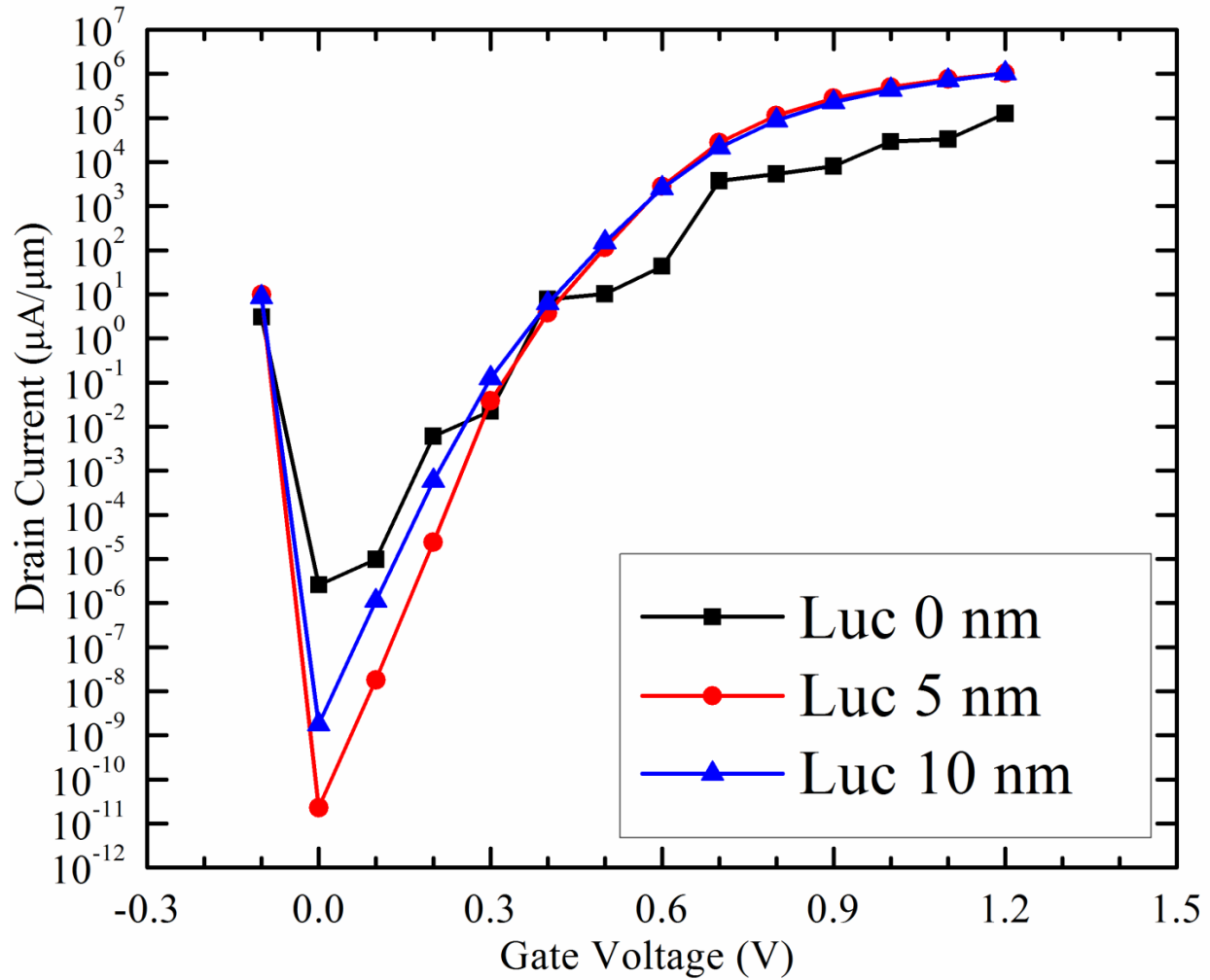

Fig. 11. $I_{DS}$–$V_{GS}$ characteristics of the InSb/InAs TFET at $V_{DS}$=0.3V with variation in undercut lengths ($L_{UC}$)

In the figure 12 we showed the variation of equivalent oxide thickness (EOT) on BG InSb/InAsTFETs. A thinner EOT gives strong coupling between InAs channel and gate field and hence, gives rise to steeper subthreshold swing. With variation in EOT from 1.9 nm to 4 nm subthreshold swing increases from 21.73 mV/decade to 31.62 mV/decade. In the thinner EOT due to strong coupling ambipolar current also increases and hence, $I_{ON}$ current slightly increases. For the InSb/InAsTFETs of given undercut lengths ($L_{UC}$) subthreshold slope depends upon the EOT. In the figure 13 we showed the variation of different Gate oxide material on BG InSb/InAsTFETs. For $Al_2O_3$ High-K Gate material subthreshold swing reducesto 20mV/decade and for $HfO_2$ High-K Gate material subthreshold swing is 18.6mV/decade.

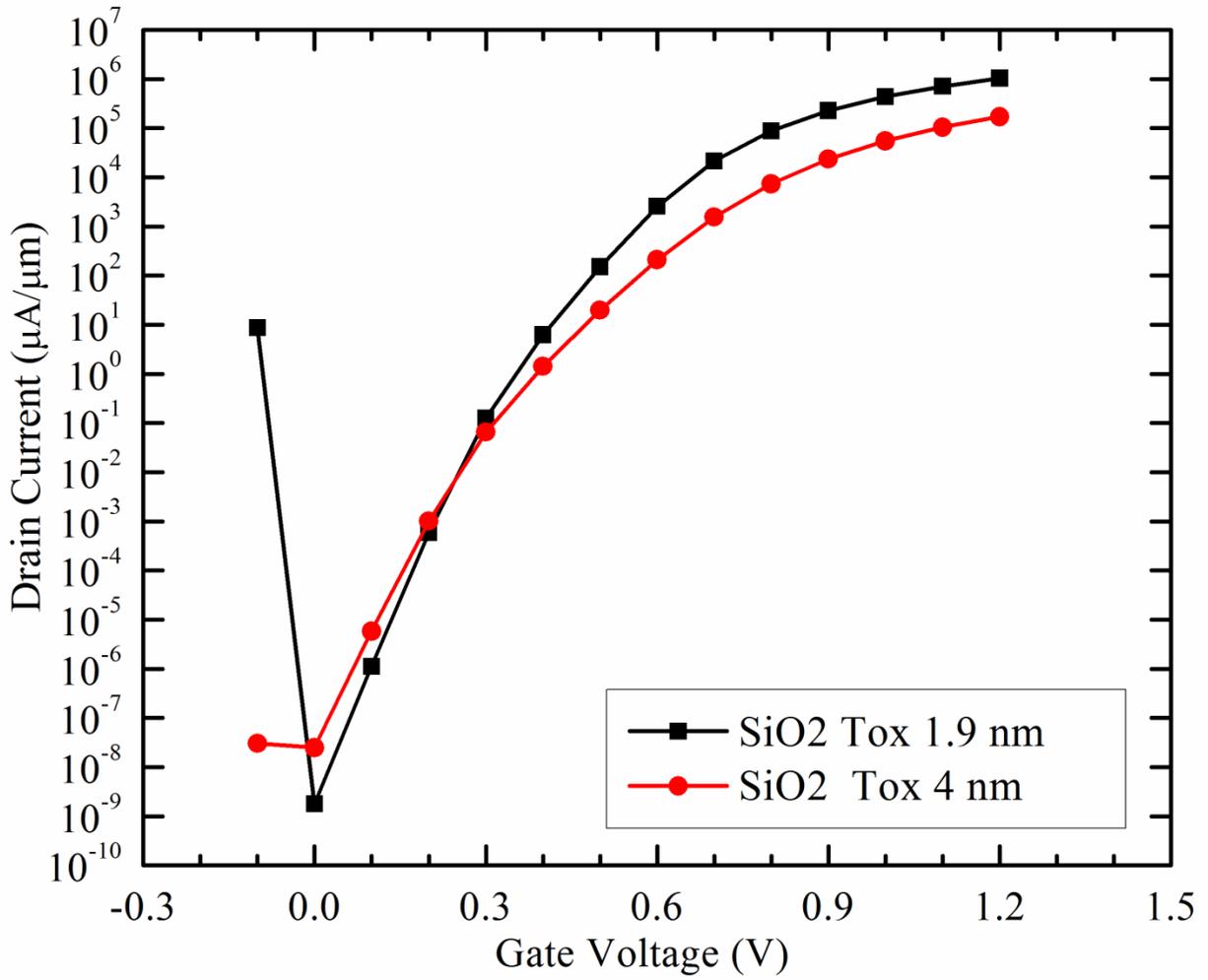

Fig. 12. $I_{DS}$–$V_{GS}$ characteristics of the InSb/InAs TFET at $V_{DS}$= 0.3V with variation inTox

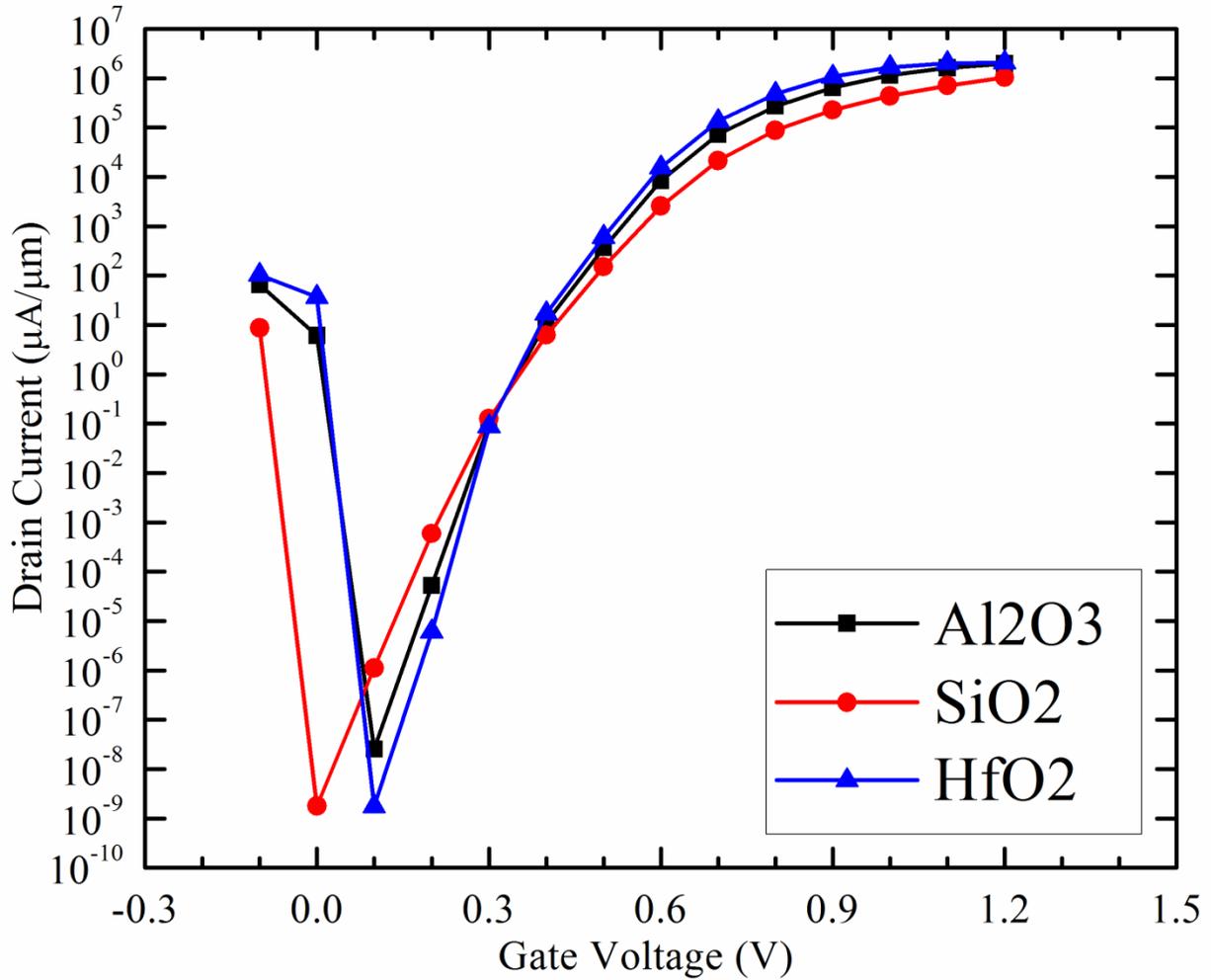

Fig. 13. $I_{DS}$–$V_{GS}$ characteristics of the InSb/InAs TFET at $V_{DS}$= 0.3V with variation in Gate Oxide Material

Infigure 14, we showed the effect of variation in gate-length ($L_G$) on BG InSb/InAsTFETs. We are varying the gate lengths ($L_G$) from 10.14 nm to 30.14 nm while holding all otherdimensionsconstant and comparing the $I_{DS}$–$V_{GS}$ transfer characteristics. Gate length ($L_G$) had a weak influence on $I_{ON}$ current. But, gate length ($L_G$) has influence on $I_{OFF}$ current and in OFF-State 2 orders of magnitude difference in $I_{OFF}$ current is obtained. However, gate length influences the subthreshold swing and as gate length decreases subthreshold swing becomes steeper. As shown in Figure 14 for gate-length of 25.14 nm $I_{OFF}$ current is $10^{-7}$μA/μm at $V_{GS}$ of 0V and for gate-length of 15.14 nm $I_{OFF}$ current is $10^{-9}$μA/μm at $V_{GS}$ of 0V.These simulation results suggest that the device performance of BG InSb/InAsTFETs is determined by gate-drain periphery and gate has good control on the center of the tunnel junction to turn off the devices.

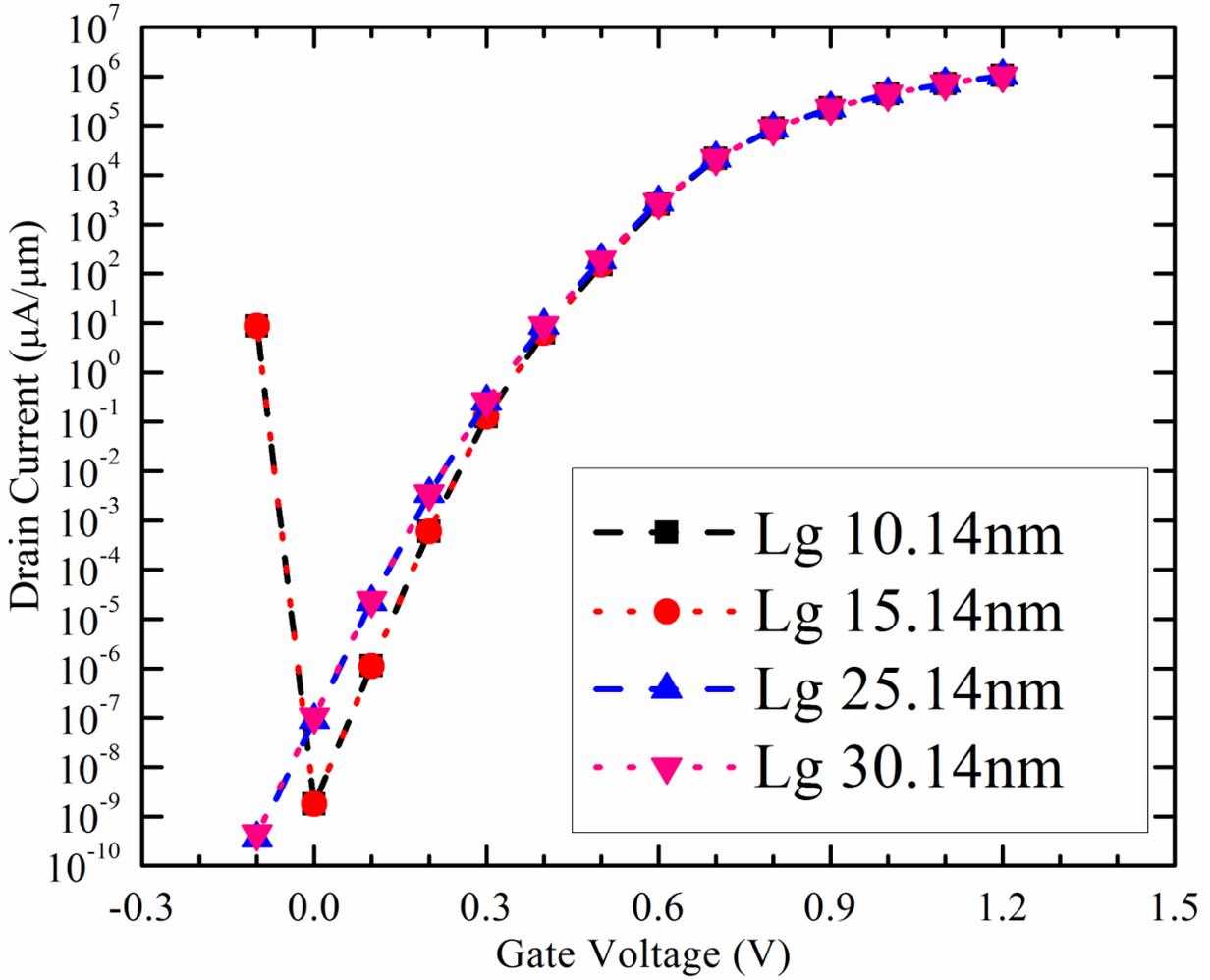

Fig. 14. $I_{DS}$–$V_{GS}$ characteristics of the InSb/InAs TFET at $V_{DS}$= 0.3V with variation in Gate Length ($L_G$)

In figure 15, we showed the effect of variation of Drain Thickness ($T_{InAs}$) in BG InSb /InAs TFETs. For the InSb /InAsTFETs of given undercut lengths ($L_{UC}$) $I_{OFF}$ current and subthreshold slope depend upon the InAs channel thickness. For larger Drain Thickness Gate has loose control on tunnel junction to shut off $I_{OFF}$ current effectively. We also observe that high $I_{ON}$ current can be achieved by higher InAs concentration but at the cost of higher value SS.

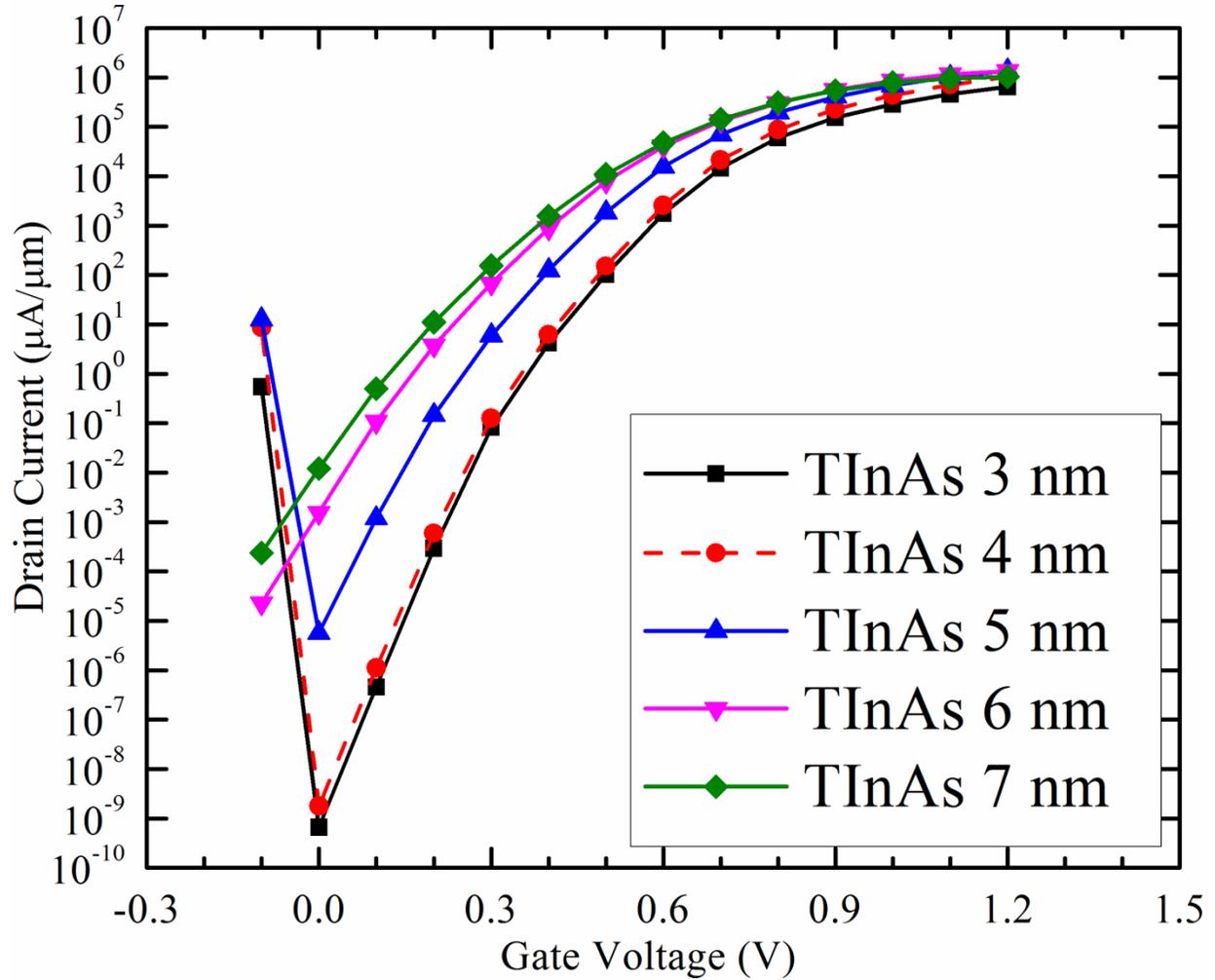

Fig.15. $I_{DS}$–$V_{GS}$ characteristics of the InSb /InAs TFET at $V_{DS}$= 0.3 V with variation in Drain Thickness ($T_{InAs}$)

CONCLUSION

L-shaped Vertical broken bandgap heterostructure InSb/InAs n-channel tunnel field-effect transistors (TFETs) of 4 nm thin channel structures with the gate lengths of 20 nm have been simulated using a 3-D, full-band, quantum mechanical simulator based on atomistic sp$^3$d$^5$s*spin-orbital coupled tight-binding method. Advantage of this geometry is that the gate electric field is in line with the tunnelling current. In this work, we used a method based on combined semiclassical electrostatic potential and NEGF calculation and applied to full quantum transport calculation of L-shaped Vertical broken bandgapInSb/InAsn-TFETs. Simulation results show that InSb/InAsn-TFETs arecapable in low-voltage operation. On applying Gate voltage $V_{GS}$ of 0.6 V along with the on state drain voltage $V_{DS}$ of 0.3 V, an $I_{ON}$ current of 1 mA/μm, an $I_{ON}/I_{OFF}$ ratio of $10^{12}$ with subthreshold swing of about 21.73 mV/decadeare obtained.We also study the effect of variation ofdifferent geometry on IV characteristics of InSb/InAsn-TFETs.These simulation results suggest that the device performance of BG InSb/InAsTFETs is determined by drain-gate periphery and gate has good control on the center of tunnelling junction to turn off the devices. In the TFETs only at the junctions high electric fields exist and current is largely determined by screened tunnelling length. Hence, gate length scaling rules of MOSFETs are not applicable for TFETs and after the leakage becomes predominant, length of intrinsic region has very negligible effect on device performances.In conclusion simulation results shownin this work demonstrate thatheterostructuresBGInSb/InAsTFET is the versatilebuilding block for future lowpower electronics and it is reasonable to experimentally investigate these TFET geometries.

## ACKNOWLEDGEMENT

The authors thank the Department of Science and Technology of the Government of India for partially funding this work.